\documentclass[10pt,  letterpaper]{IEEEtran}
\usepackage{todonotes}
\usepackage[nocompress]{cite}
\usepackage{url}
\usepackage{enumitem}
\usepackage{epstopdf}
\PrependGraphicsExtensions{.tif, .tiff}
\usepackage{graphicx}
\usepackage{caption}
\usepackage{amsmath}
\usepackage{tabu}
\usepackage{multirow}
\usepackage{adjustbox}
\usepackage{mathptmx}
\usepackage{amsfonts}
\usepackage{algpseudocode}
\usepackage{soul}
\usepackage[bookmarks=false]{hyperref}
\usepackage{footnote}
\usepackage{lipsum}
\usepackage[ruled,vlined, linesnumbered]{algorithm2e}
\usepackage{subcaption}
\usepackage{listings}
\usepackage{booktabs}
\usepackage{tikz}
\usetikzlibrary{positioning}
\soulregister\cite7
\soulregister\ref7
\soulregister\pageref7
\soulregister\secref7
\soulregister\tabref7

\newcommand*\circled[1]{\tikz[baseline=(char.base)]{
            \node[shape=circle,draw,inner sep=0.75pt, text=white,fill=black] (char) {#1};}}

\definecolor{ao}{rgb}{0.0, 0.5, 0.0}
\definecolor{db}{rgb}{0.2, 0.2, 0.6}
\definecolor{cadmiumgreen}{rgb}{0.0, 0.42, 0.24}

\newcommand{\tabref}[1]{Table~\ref{#1}}

\newcommand{\secref}[1]{Section~\ref{#1}}

\SetAlFnt{\footnotesize}

\begin{document}
\title{FedLess: Secure and Scalable Federated Learning Using Serverless Computing}
\author{\IEEEauthorblockN{Andreas Grafberger\IEEEauthorrefmark{1}, Mohak Chadha\IEEEauthorrefmark{1}, Anshul Jindal\IEEEauthorrefmark{1}, Jianfeng Gu\IEEEauthorrefmark{1}, Michael Gerndt\IEEEauthorrefmark{1}\\}
\IEEEauthorblockA{\IEEEauthorrefmark{1}Chair of Computer Architecture and Parallel Systems, Technische Universit{\"a}t M{\"u}nchen \\
Garching (near Munich), Germany \\} 
Email: \{andreas.grafberger, mohak.chadha, anshul.jindal, jianfeng.gu\}@tum.de, gerndt@in.tum.de}

\maketitle


\begin{abstract}

The traditional cloud-centric approach for Deep Learning (DL) requires training data to be collected and processed at a central server which is often challenging in privacy-sensitive domains like healthcare. Towards this, a new learning paradigm called Federated Learning (FL) has been proposed that brings the potential of DL to these domains while addressing privacy and data ownership issues. FL enables remote clients to learn a shared ML model while keeping the data local. However, conventional FL systems face several challenges such as scalability, complex infrastructure management, and wasted compute and incurred costs due to idle clients. These challenges of FL systems closely align with the core problems that serverless computing and Function-as-a-Service (FaaS) platforms aim to solve. These include rapid scalability, no infrastructure management, automatic scaling to zero for idle clients, and a pay-per-use billing model. To this end, we present a novel system and framework for serverless FL, called \emph{FedLess}. Our system supports multiple commercial and self-hosted FaaS providers and can be deployed in the cloud, on-premise in institutional data centers, and on edge devices. To the best of our knowledge, we are the first to enable FL across a large fabric of heterogeneous FaaS providers while providing important features like security and Differential Privacy. We demonstrate with comprehensive experiments that the successful training of DNNs for different tasks across up to 200 client functions and more is easily possible using our system. Furthermore, we demonstrate the practical viability of our methodology by comparing it against a traditional FL system and show that it can be cheaper and more resource-efficient.

\end{abstract}


\IEEEpeerreviewmaketitle
\thispagestyle{empty}

\section{Introduction}
\label{sec:intro}

Heterogeneous remote edge devices or siloed data centers such as mobile phones or hospitals generate manifolds of data each day~\cite{chiang2016fog}. Using deep learning, the data generated by these devices and institutions can be used to enable smart applications~\cite{NvidiaClara2020}. In the conventional deep learning~\cite{lecun2015deep} approach, all the raw training data is collected and stored on a central server. However, increasing privacy concerns of data holders and recent legislation on data protection and privacy such as the European General Data Protection Regulation (GDPR)~\cite{EUdataregulations2018}  prevent the transmission of data to a centralized location, thus making it impossible to train DNNs on data stored and processed in different locations. Towards this, a new paradigm for distributed ML training called Federated Learning (FL)~\cite{mcmahan2017communication} has been introduced. FL enables the collaborative training of ML models and addresses the fundamental problems of privacy and ownership of data. In FL, remote clients learn a shared model by optimizing its parameters on their local data and sending back the updated parameters. These local model updates are then aggregated to form the new, updated shared model. Clients in FL can be mobile, edge devices, institutions operating their own data centers, or Virtual Machines managed by Infrastructure-as-a-Service (IaaS) providers~\cite{Kairouz2019}. In this paper, we argue that both components in a traditional FL system, i.e., clients and server, can immensely benefit from a new computing paradigm called serverless computing.  

In serverless computing, developers do not have to manage infrastructure themselves but completely hand over this responsibility to a Function-as-a-Service Platform~\cite{cncf-serverless-whitepaper}. Several open-source and commercial FaaS platforms such as OpenWhisk~\cite{openwhisk}, OpenFaaS~\cite{openfaas}, AWS Lambda~\cite{aws-lambda}, and Google Cloud Functions (GCF)~\cite{gcloud-functions} are currently available. Applications are developed as small units of code, called functions that are independently packaged and uploaded to a FaaS platform and executed on event triggers such as HTTP requests. In the context of serverless, FL clients are functions deployed on a FaaS platform that are capable of calculating the shared model updates. A connected group of different FaaS platforms is referred to as a FaaS fabric~\cite{Chadha2020}.

On the client-side, using FaaS technologies can lead to improvements in resource-efficiency and cost. For instance, a common practice in FL is to only select a small fraction of available clients for each training round. Increasing client parallelism is only beneficial up to a certain point with diminishing returns when more clients are included in each training round~\cite{bonawitz2019federated,mcmahan2017communication}. As a result, most clients wait until they are selected in a later round and stay idle, leading to wastage of resources and incurred costs if the clients use an IaaS provider. The idle times are further enhanced due to the potential heterogeneity in clients’ computing capabilities or local dataset size (\S\ref{sec:flback}). On the FL server-side, one common challenge is to efficiently handle sudden bursts of compute-intensive workloads whenever clients report their local updates for aggregation~\cite{bonawitz2019federated}.  In addition to wasted compute or scalability challenges, requiring all data holders to manage the complex infrastructure for their clients can be an enormous burden. These challenges of current FL systems closely align with the core problems that serverless technologies and FaaS platforms aim to solve. These include rapid scalability during request bursts, automatic scaling to zero when resources are unused, and an attractive pricing and development model~\cite{cncf-serverless-whitepaper}.

Towards efficient, scalable, and enabling ease-of-use for FL, our key contributions are:

\begin{itemize}
    \item We present a novel system and framework called \emph{FedLess}\footnote{https://github.com/andreas-grafberger/fedless} to perform FL on a heterogeneous fabric of FaaS platforms with support for arbitrary DNN models using Tensorflow~\cite{tensorflow}, proper authentication and authorization of clients, and a privacy-protecting training mechanism.   
    
    \item \emph{FedLess} supports four major commercial FaaS platforms, i.e., AWS Lambda, GCF, Azure Functions~\cite{azure-functions} and IBM Cloud Functions~\cite{ibm-cloud-functions}, and two major open-source platforms, i.e., OpenWhisk~\cite{openwhisk} and OpenFaaS~\cite{openfaas}. We provide examples and helper functions to run FL client functions on all these platforms.
    \item We demonstrate with extensive experiments the features and scalability of \emph{FedLess} on three popular FL benchmark datasets using different DNN models.
    \item We compare \emph{FedLess} with an optimized reimplementation of an existing FaaS based FL system~\cite{Chadha2020} and demonstrate that our system is significantly faster.
   \item We practically demonstrate the viability of \emph{FedLess} by comparing it with a traditional IaaS based FL system using Flower~\cite{flower} with respect to performance and cost.
    
\end{itemize}



The rest of this paper is structured as follows. \S\ref{sec:background} provides a background on FaaS and FL. In \S\ref{sec:related_work}, the previous work on serverless ML, FL, and other FL frameworks are described. \S\ref{sec:fedless} describes our system and framework \emph{FedLess}. In \S\ref{sec:setup}, our experimental setup is described, while \S\ref{sec:results} presents our experimental results. Finally, \S\ref{sec:futurework} concludes the paper and presents an outlook.


\begin{figure}[t] 
    \centering
    \includegraphics[width=0.75\columnwidth]{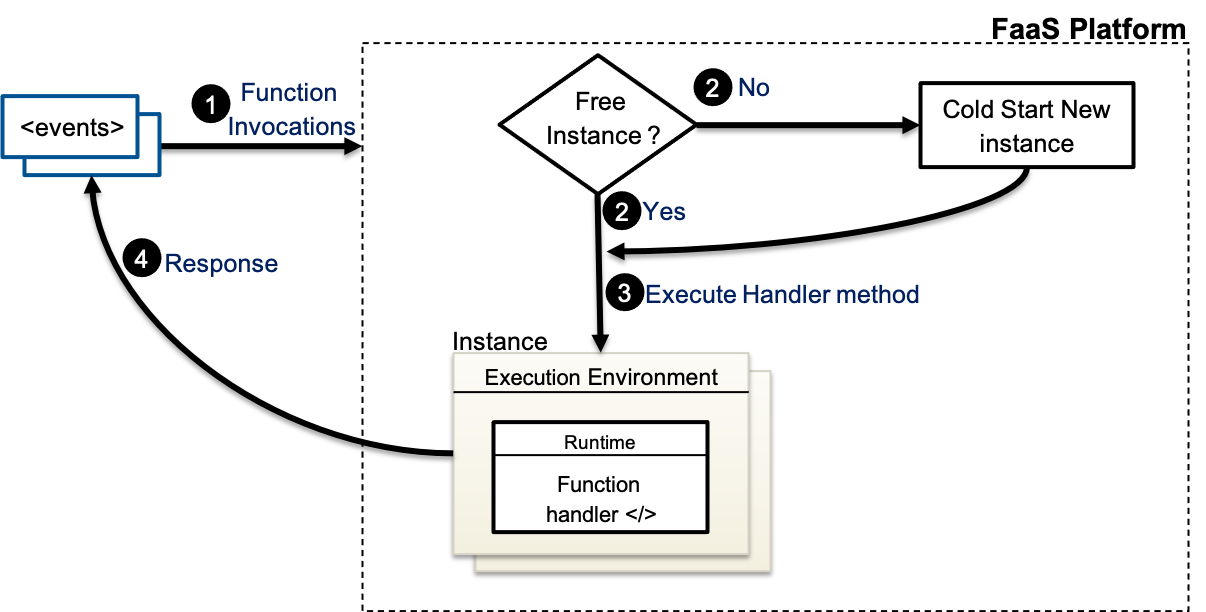}
    \caption{Workflow demonstrating functioning of the FaaS paradigm.}
    \label{fig:faasworkflow}
\end{figure}

\section{Background}
\label{sec:background}

\subsection{Function-as-a-Service (FaaS)}
\label{sec:bfaas}
Due to its simplicity, client-friendly cost model, and automatic scaling, FaaS is emerging as a preferred cloud-based programming paradigm. It has gained popularity and widespread adoption in different application domains such as linear algebra~\cite{serverlesslin} and heterogeneous computing~\cite{jindal2021function, postericdcs, courier}. An overview of the functioning of the FaaS paradigm is shown in Figure 1. In FaaS, the user implements fine-grained functions that are executed in response to event triggers or HTTP requests (\circled{1}). On function invocation, the FaaS platform is responsible for providing resources to the function and its isolation in ephemeral, stateless containers. These containers are commonly referred to as function instances~\cite{wang2018peeking, fncapcatior} and contain the function's language-specific runtime environment. For commercial FaaS platforms, the function instances are launched on the FaaS platforms' traditional Infrastructure-as-a-Service Virtual Machines as described in~\cite{chadha2021architecturespecific, kiener2021demystifying}. The runtime is responsible for relaying the invocation events, context information, and responses between the FaaS platform and the function. On the first invocation of a function, the FaaS platform creates a new function instance, i.e., cold start~\cite{agile} and runs its handler method to process the event (\circled{2}-\circled{3}). When the handler returns a response or exits (\circled{4}), the function instance remains active for a specific duration to handle successive events. The FaaS platform can create concurrent function instances to handle multiple events or requests on-demand. When the number of requests decreases, the FaaS platform automatically scales down the number of active function instances. For commercial FaaS platforms, users are billed based on the execution time of the functions measured in 100ms or 1ms intervals, i.e., pay-per-use. Moreover, most commercial FaaS platforms limit the maximum execution time and memory a function can have. For instance, with AWS Lambda, these limits are $15$ minutes and $10$GB, respectively. However, due to the rapid development in FaaS, these limits might disappear or be relaxed soon~\cite{Fox2017}.


\begin{figure}[t] 
    \centering
    \includegraphics[width=0.75\columnwidth]{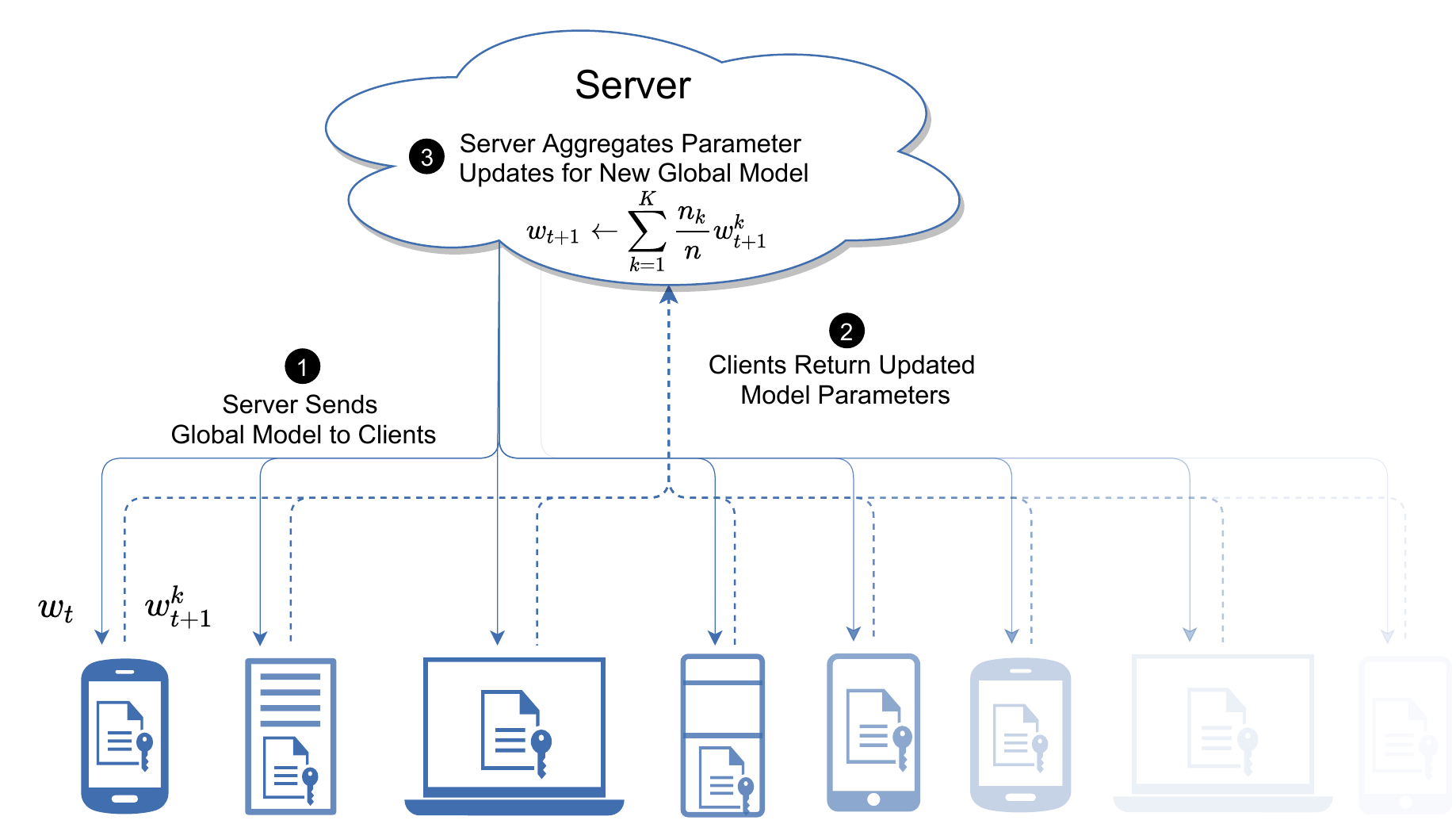}
    \caption{General synchronous training workflow in FL.}
    \label{fig:fl-example-system}
\end{figure}



\subsection{Federated Learning (FL)}
\label{sec:flback}
FL describes a new learning paradigm for data holders to collaboratively train an ML model while keeping their data private. A typical FL system has two main entities, i.e., \emph{clients} and a \emph{central server}. Depending on the FL category, i.e., \emph{cross-device} or \emph{cross-silo}, the clients in FL can range from mobile, edge devices to machines running in cloud data centers~\cite{Kairouz2019}. In \emph{cross-device} setting, the number of clients can be significantly large with limited network and compute capabilities, while in \emph{cross-silo} setting, there are few clients, all with abundant computing resources. Figure~\ref{fig:fl-example-system} describes the general synchronous training workflow for both FL categories. At the start of each round (\circled{1}), the server sends the current (at the start randomly initialized) global model to all the clients. Following this, each client improves the shared model by optimizing it on its local dataset and sends back only the updated model parameters (\circled{2}). Finally, the local model updates from all participating clients are collected and aggregated (\circled{3}). The server usually does not possess its own dataset and is primarily responsible for organizing the training and deciding which clients contribute in a new round. The default implementation in \emph{FedLess} uses the \texttt{FedAvg}~\cite{mcmahan2017communication} algorithm for aggregating the model parameters. In contrast to traditional distributed ML, the training data present on clients in FL is \emph{non-IID}, i.e., one client's dataset is not representative of the full data distribution across all clients, and \emph{unbalanced}, i.e., a small number of clients can have large local datasets while other clients can contain only a few records.

\section{Related Work}
\label{sec:related_work}
\textbf{Serverless Machine Learning}. 
The majority of the previous works~\cite{Wang2019a, Carreira2019, Jiang2021} in this domain have focused on distributed ML training using FaaS functions. \emph{Siren}~\cite{Wang2019a},  allows users to train ML models in the cloud using fine-grained functions, thereby removing the burden of non-trivial cluster provisioning and management from the developers. Furthermore, it enables asynchronous distributed training and features an algorithm based on Deep Reinforcement Learning that continuously tunes the number of functions and their memory to optimize performance and cost. In contrast to \emph{Siren}, \emph{Cirrus}~\cite{Carreira2019} supports complete end-to-end ML workflows including data preprocessing and hyperparameter tuning. It provides a lightweight worker runtime for the cloud functions that support various ML models. Jiang et al.~\cite{Jiang2021} analyze the cost-performance trade-offs between Infrastructure-as-a-Service (IaaS) and FaaS for distributed ML. Towards this, they developed \emph{LambdaML}, which supports different distributed ML variants such as synchronous vs. asynchronous training, purely FaaS-based, or a hybrid (FaaS/IaaS) training approach. In contrast to previous works, \emph{FedLess} differs in several ways. First, although FL and distributed ML share some similarities, they have key fundamental differences. For instance, data in FL is private to the client and not accessible by the central server. Furthermore, in distributed ML, workers are rarely idle, whereas, in FL, usually not all workers participate in a round of training. To this end, we designed \emph{FedLess} from the ground up, keeping FL in mind. Second, none of the tools mentioned above are vendor-agnostic and support only one FaaS platform, i.e., AWS Lambda. In contrast, \emph{FedLess} supports clients distributed across all major commercial and open-source FaaS platforms. Third, none of the tools consider security aspects for function invocations and model training. On the other hand, \emph{FedLess} supports cloud-agnostic function authentication/authorization. Finally, \emph{FedLess} is built on top of Tensorflow~\cite{abadi2016tensorflow}, which allows easy integration and usage for developers. In contrast, \emph{Siren} features a runtime based on a pruned version of MXNet~\cite{chen2015mxnet}, while \emph{Cirrus} does not support any DL framework.


\textbf{Federated Learning Frameworks}.
There exists several open-source frameworks for FL such as \emph{PySyft}~\cite{pysyft}, \emph{Tensorflow Federated} (TFF)~\cite{Bonawitz2020}, \emph{Flower}~\cite{flower}, \emph{FedML}~\cite{He2020}, \emph{PaddleFL}~\cite{paddle}, and \emph{Fate}~\cite{fate}. Although many of these frameworks provide implementations for different FL algorithms and privacy mechanisms by default, none of them have been developed to be used in a serverless context. Moreover, they require all participating clients to rely on specific communication protocols such as \texttt{gRPC}. As a result, for maximum flexibility and to optimize our implementation for the serverless use-case, we do not use them.


\textbf{Serverless Federated Learning}. To the best of our knowledge, we are the only ones to have tried FaaS for FL. In our previous work~\cite{Chadha2020}, we presented \emph{FedKeeper}, a tool to facilitate FL of ML models across a fabric of heterogeneous FaaS platforms. However, \emph{FedKeeper} lacks crucial features that are required by FL systems in practice. For instance, it only supports relatively small models, does not provide support for security or a privacy protection mechanism, and is not optimized for performance. In this paper, we present a new system, \emph{FedLess} that addresses the drawbacks present in \emph{FedKeeper}. Moreover, \emph{FedLess} is designed to be modular and can be easily extended to support additional data sources, models, and optimization algorithms.

\section{Fedless}
\label{sec:fedless}

\emph{FedLess} is a system and framework to perform FL on a heterogeneous fabric of FaaS platforms. In this section, we first describe the overall architecture of \emph{FedLess}. Following this, we present an overview of its implementation, describe its security features, the mechanism for privacy-preserving training of clients, and performance optimizations that improve scalability and reliability in detail. Finally, we describe the workflow for training clients using \emph{FedLess}.



\begin{figure}[t]
    \centering
    \includegraphics[width=1.0\columnwidth]{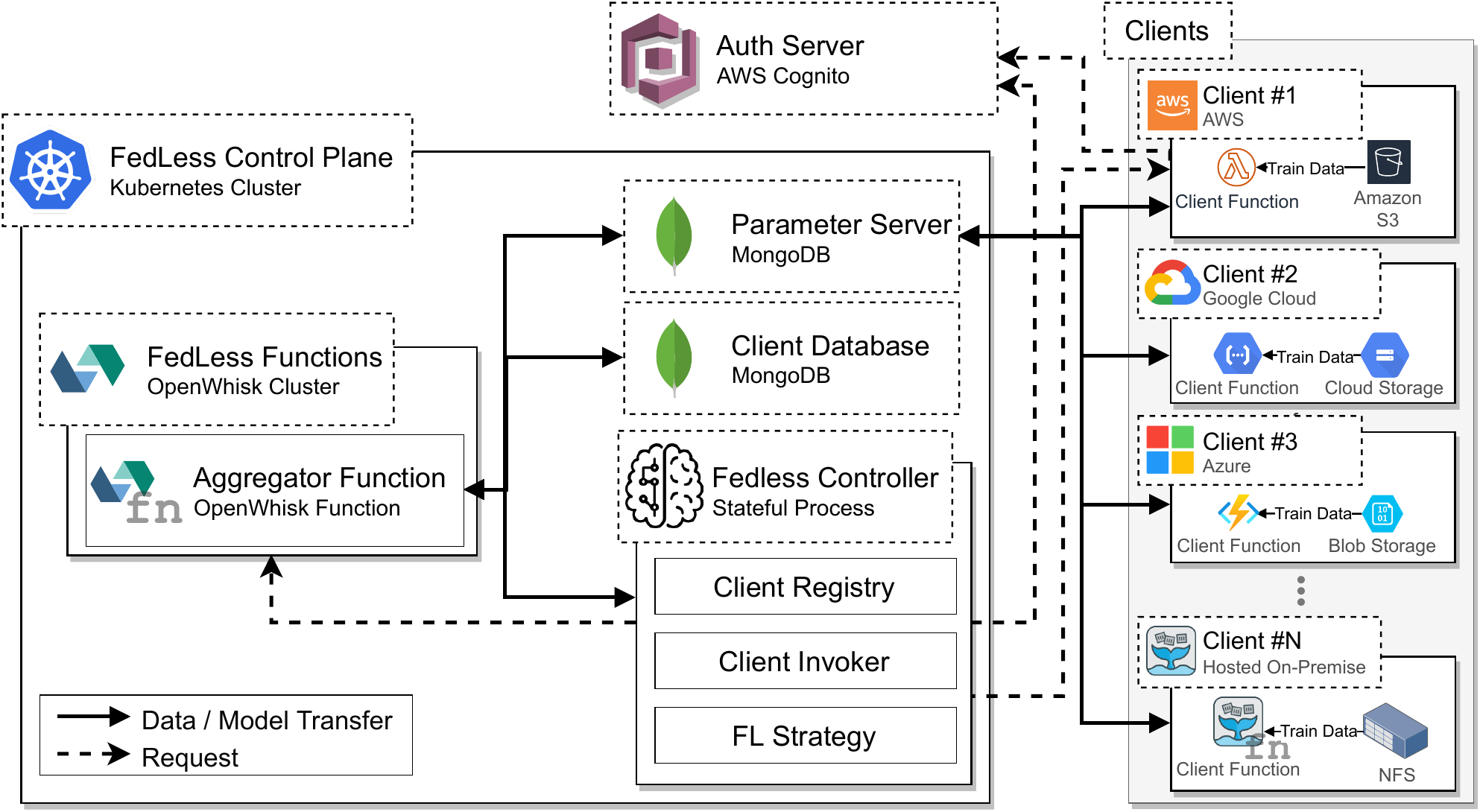}
    \caption{FedLess System Architecture.}
    \label{fig:fedlesssysarch}
\end{figure}

\subsection{System Design}
\label{sec:sysdesign}

Figure~\ref{fig:fedlesssysarch} gives an overview of the different components of \emph{FedLess}. In the serverless context,  we refer to an FL \emph{client} as the associated \emph{client function}, i.e., the function deployed on a FaaS platform that calculates the local model updates. A \emph{client administrator} is the person who is responsible for managing and deploying the function and is usually also the data holder. The FL training process and the registration of clients are managed by the FL \emph{administrator} (FL admin).

\emph{FedLess} supports both \emph{cross-silo} and \emph{cross-device} FL use-cases (\S\ref{sec:flback}). In the \emph{cross-silo} setting, client functions would belong to a few institutions and either run in on-premise data centers with self-hosted FaaS platforms or on the cloud. In the latter, client functions would run on edge devices with lightweight FaaS platforms such as \emph{faasd} (OpenFaaS)~\cite{faasd}. All clients are externally deployed functions that are callable through an HTTP interface. By default, \emph{FedLess} supports four commercial and two open-source FaaS platforms. We provide scripts for deploying client functions on these FaaS platforms. Apart from setting up optional parameters, such as request limits, no further action is required from the user.

The central component in \emph{FedLess} is the control plane which comprises of the \emph{Parameter Server}, the \emph{Client Database}, an OpenWhisk cluster to run the \emph{Aggregator Function}, and the \emph{FedLess Controller}. The controller comprises of a \emph{Client Registry}, \emph{Client Invoker}, and the \emph{FL Strategy}. The default FL strategy implemented in our system is the \texttt{FedAvg} algorithm~\cite{mcmahan2017communication}. Due to its ease-of-use, reliability, replication support, and security features, we chose MongoDB~\cite{mongodb} as our parameter server. Moreover, we make use of MongoDB's \emph{GridFS} specification to support models larger than the document size limit, i.e., \texttt{16MB}. The global model architecture and its hyperparameters are also stored in the parameter server. The \emph{Client Database} is also a MongoDB instance that serves as a persistent registry for the client functions, their hyperparameters, and dataset locations, along with any additional information supplied by the client administrator during sign-up. The \emph{Controller} is a lightweight process that manages and monitors the entire training cycle. It is responsible for selecting the clients that participate in each round, invoking them and the aggregator. The client registry component in the controller interacts with the client database for managing the clients. The aggregator is a FaaS function that is invoked by the controller after all clients that participated in a training round have finished. It aggregates the updated client model parameters according to the FL strategy.

New FL strategies other than \texttt{FedAvg} can easily be implemented by modifying how the controller samples clients for each round and writing a custom aggregator function, taking the one provided as a blueprint. Having the parameter server physically close to the aggregation process is a crucial advantage of our approach due to minimized networking overhead, as is also presented in~\cite{Jiang2021}. Although we deploy the aggregator function using OpenWhisk as shown in Fig~\ref{fig:fedlesssysarch}, it can be easily deployed using another FaaS platform. The \emph{Auth Server} is an external entity responsible for authentication and is described in detail in \S\ref{sec:authenticationauth}.



\subsection{Implementation Overview}
\label{sec:imploverview}
The \emph{FedLess} framework was implemented using Python3 and also includes a command line tool to orchestrate the entire FL training process. It relies on Tensorflow~\cite{tensorflow} and Keras~\cite{keras} to support arbitrary DNN models. Our framework provides:

\begin{itemize}
\item Low-level, easily exchangeable implementations for serialization methods, helper functions for tasks related to security and general optimizations, as well as classes for easy database access and interaction with the parameter server.
\item Functions and decorators to abstract away differences of supported FaaS platforms, in turn, unifying the implementation of client functions.
\item Extensive example implementations of client and aggregation functions that serve as easily modifiable blueprints.
\item Custom data types and models using the \texttt{Pydantic}~\cite{pydantic} library, which ensures that all clients and the FL server rely on the same data formats and schemata.
\end{itemize}




\subsection{Security}
\label{sec:syssecurtiy}
Since the clients in FL belong to separate institutions and networks and have to expose their functionality to the public internet to be accessible by the FL server, it is significantly important that only authenticated and authorized entities can invoke client functions. The security features in \emph{FedLess} are described in the following four aspects.

\subsubsection{Function Ownership}
\label{sec:funcownership}
In \emph{FedLess}, all participating institutions and users that manage one or multiple clients are responsible for deploying the client functions themselves. This is because data holders have complete control over what happens with their data. Moreover, this approach provides greater flexibility for institutions to adjust their complete end-to-end local training workflows. In addition, full oversight on security and access control is essential, especially for domains such as healthcare~\cite{Xu2019a} with special legal regulations for the data. Clients deployed on different FaaS platforms can directly use the authentication system provided by \emph{FedLess} (\S\ref{sec:authenticationauth}). However, since data holders are free to use different FaaS platforms, they can require additional, platform-specific authentication secrets. A data holder could, for example, deploy a client function with OpenWhisk and specify an API token that has to be supplied by \emph{FedLess} with every request. Our system supports additional API tokens in client invocation requests. We provide several examples for writing client functions using our framework for multiple FaaS platforms.



\subsubsection{Authentication and Authorization}
\label{sec:authenticationauth}
In a centralized FL system, like \emph{FedLess}, multiple security interests come together. For instance, data holders require that only the central FL server they integrate with can call their client functions. This means that the central server has to authenticate itself and possess proper authorization. At the same time, before being added to the list of valid client functions for an FL training session, data holders have to identify themselves to the FL server and provide appropriate proof. As a result, we need to support authentication, authorization, and integration with external identity providers through protocols like \emph{OAuth 2.0}~\cite{RFC6749} and \emph{SAML 2.0}~\cite{saml-2}. Integration with external identity providers enables easy and direct support for identity providers already used by academic institutions and enterprises. Towards this, we use \emph{AWS Cognito}~\cite{aws-cognito}. It is a service that provides all the required features, has been used in large-scale production systems, and has a convenient pricing structure that scales with the size of the system. Furthermore, it does not restrict \emph{FedLess} to a specific cloud provider's ecosystem since the interaction with Cognito is the same for all the clients.

The authentication and authorization system is based on \emph{JSON Web Tokens}~\cite{RFC7519} (JWTs) attached to all HTTP requests between the FL server and the clients. Client administrators manually register through a web interface with a Cognito User Pool, offering sign-up via external identity providers. Once they submit their registration request, a central FL server administrator checks the validity of the registration information and confirms it. After the registration process, all necessary information for the client function to check the authenticity and permissions of requests by the FL server is then communicated to the client administrator via a web interface. This includes the public key of the key-pair used to sign the authentication headers, the FL server’s client-id, and required authorization scopes. At the start of each training round, the FL server uses its credentials to fetch a token from Cognito with sufficient privileges to invoke all client functions. For each new request, all clients first check if the token supplied with the request by the FL server was signed by the trusted Cognito User Pool and has the proper authorization.


\subsubsection{Parameter Server Access}
\label{sec:parameterserveraccess}
The different client functions need to access the parameter server without compromising security and privacy. For instance, one client function should not be able to read the result of another client or modify the global parameters. To this end, the FL server creates temporary users in the parameter server for each client function that can only read the shared global model and write back results without access to other clients. In every request to a client, the FL server attaches the credentials with which the client can retrieve the global model and write back its results to the parameter server. Furthermore, because the credentials given to the client function belong to a custom user, the access to the parameter server could be easily revoked for individual client functions. Our parameter server, i.e., MongoDB (\S\ref{sec:sysdesign}) allows the definition of such custom users and role policies. 




\subsubsection{General Security Features}
\label{sec:gensecfeatures}
The Open Web Application Security Project~\cite{owasp-top-ten} represents the top ten issues that are widely accepted to be most common and critical in software applications. While most security concerns depend on the system and function implementation details and lie in the hands of the data holders, we counter some of the most common vulnerabilities with our system design and implementation in \emph{FedLess}. First, since all function requests use HTTP, we can easily enforce all communication to use Transport Layer Security (TLS) to encrypt the exchanged data. Second, to protect against insecure serialization and injection attacks, we advise data holders to use the model schemata provided in our framework implemented using the \texttt{Pydantic}~\cite{pydantic} library. Third, we provide examples and demonstrate how to isolate all authentication and authorization by using separate functions as suggested by~\cite{Michener2020}.

\subsection{Privacy}
\label{sec:sysprivacy}
The two main concerns in an FL system in which not all participants can be trusted are \emph{poisoning} and \emph{inference} attacks~\cite{Lyu2020, Enthoven2021}. As a result, several strategies to protect the privacy of clients for FL have been proposed. These include \emph{Secure Multiparty Computation} (SMC)~\cite{Truex2018}, \emph{Homomorphic Encryption} (HE)~\cite{Phong2018}, and \emph{Differential Privacy} (DP)~\cite{Dwork2008, Dwork2013}. While SMC and HE involve encrypting the client updates, DP involves adding Gaussian noise to them. Due to its reliable privacy guarantees at the cost of degraded prediction accuracy, it is a common practice to combine DP with SMC or HE methods~\cite{Kairouz2019,Truex2018}. This is because due to encryption, less noise would have to be added to the client updates. However, a drawback of these encryption-based approaches is that they require a large amount of memory and have significant computational and networking costs~\cite{Phong2018,Zhang2020,Zhu2020a}. This makes them unsuitable for most FL applications~\cite{Bonawitz2017}. For FaaS, using these methods leads to function timeouts due to fixed limits (\S\ref{sec:bfaas}). As a result, we do not employ secure aggregation in our system. By default, \emph{FedLess} supports \emph{Local Differential Privacy}~\cite{Mothukuri2021,Kim2021,Wei2020} (LDP). LDP is a form of  \emph{record-level} privacy~\cite{Melis2019} in which the client functions add noise to their parameters before uploading them to the parameter server. 

We use the \emph{Tensorflow Privacy}~\cite{McMahan2018,tf-privacy} library that provides DP versions of popular optimization algorithms that are also compatible with Keras. A minor caveat of using LDP with ephemeral serverless functions is keeping track of past invocations to a client function through an external storage system. Subsequent invocations of the client function degrade the privacy guarantees~\cite{Kim2021}. As a result, data holders would want to prohibit requests by the FL server after a specific privacy budget has been exhausted.  For this, they would need to store their state between requests. All of this, combined with our reference implementation of using  DP optimizers with client functions, makes it easy for data holders to slightly modify their client function code to benefit from DP and provide proven privacy guarantees for their user data. Investigating the use of other secure aggregation methods such as  \emph{Additive Masking}~\cite{Bonawitz2017} which require complex interaction between clients, is of our interest in the future.

\subsection{Performance Optimizations}
\subsubsection{Global namespace caching} 
\label{sec:globalcache}
Due to the ephemeral, stateless nature of FaaS functions, they need to load and preprocess the dataset, compile and instantiate the model on every invocation. This overhead becomes significantly large for large local datasets and models. To mitigate this, we exploit that all cloud providers and FaaS frameworks retain global variables in the function instances between invocations. Towards this, we implement a custom Python caching decorator that, after the first function invocation, stores the output of the Python function in an in-memory LRU cache. All subsequent calls to the function return the cached result. Note that the cached result is only available until the duration of the function instance (\S\ref{sec:bfaas}).

\subsubsection{Running average model aggregation} 
\label{sec:rama}
In a naive implementation of \texttt{FedAvg}, the aggregator function has to load all model updates in memory for the aggregation. However, due to the limitations on the amount of memory configurable for FaaS functions (\S\ref{sec:bfaas}) on most commercial FaaS providers, this approach is not feasible. To this end, we implement a running average mode in the aggregator function. This is done by loading the client updates from the parameter server only in small batches. After a batch is loaded, we use it to update the current average global model parameters and free up the memory occupied by the processed batch.

\subsubsection{Federated Evaluation}
\label{sec:federatedeval}
The aggregator function can evaluate the updated global model after each round if a global test exists. However, in FL, since this is rare, we also support client-side evaluation as in~\cite{Caldas2018}. Each client can be invoked independently to run the global model on a local test set and return the evaluation metrics. The metrics can then be aggregated by the controller. Moreover, independent selection of clients for training and evaluation between rounds minimizes function cold starts.

\begin{figure}[t]
    \centering
    \includegraphics[width=1.0\columnwidth]{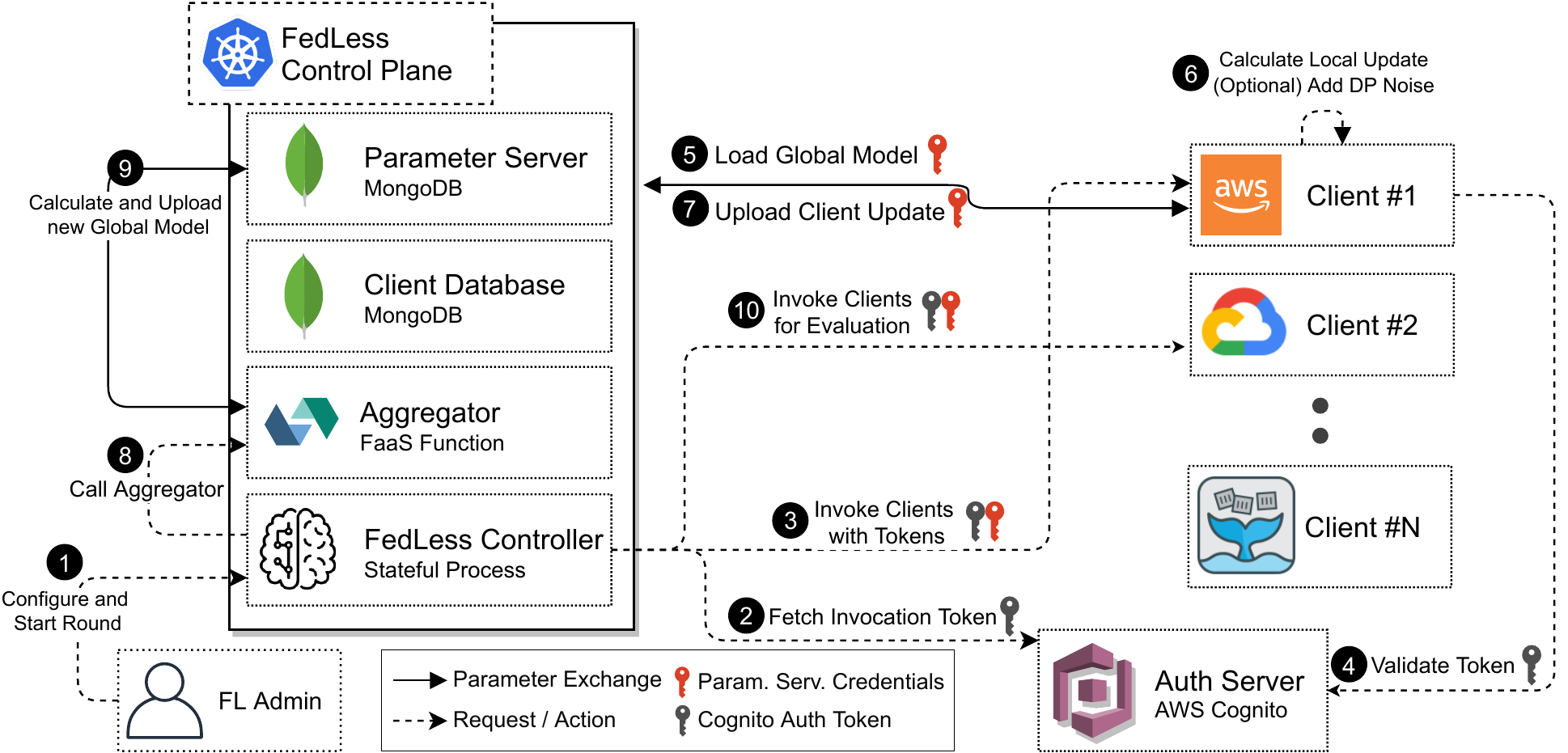}
    \caption{FedLess workflow for a complete FL round.} 
    \label{fig:fedless-workflow}
\end{figure}

\subsection{Putting It All Together}
\label{sec:sysworkflow}
The workflow for training multiple clients using \emph{FedLess} in a single FL round is presented in Figure~\ref{fig:fedless-workflow}. First, the FL admin (\S\ref{sec:sysdesign}) selects the model to be trained, the registered client functions that will participate in the training, and hyperparameters such as the number of clients per round (\circled{1}). Following this, the training is started by the FL admin. The \emph{FedLess} controller first requests a new invocation token from the Auth Server using the credentials configured by the FL admin (\circled{2}). Using this token along with the credentials to access the parameter server, the controller invokes the clients it selected for this round (\circled{3}). The clients involved in an FL training round are selected randomly by the controller. To ensure that the invocation is valid, the clients validate the signature and authorization of the token using the public key from the Auth Server (\circled{4}). The clients then use the supplied parameter server credentials to load the latest global model (\circled{5}). Following this, the clients load their local datasets and perform the local training, optionally using LDP (\circled{6}). Once the clients have finished training, they again use their credentials to upload their parameters to the parameter server (\circled{7}). The \emph{FedLess} controller waits until all clients are either finished, reached a configured timeout, or failed. It then starts the model aggregation by invoking the aggregator function (\circled{8}). The aggregator loads the client results from the parameter server, aggregates the parameters, and stores the new global model (\circled{9}). Finally, the controller starts the evaluation (\circled{10}). If a global test set exists, then the evaluation was already done in the aggregator. If not, a new selection of clients is invoked for evaluation in which they load the updated model and evaluate on their test set (\S\ref{sec:federatedeval}). The controller aggregates the returned metrics and resumes the training process from (\circled{2}) if a configured accuracy threshold is not reached.



\section{Experimental Setup}
\label{sec:setup}
In this section, we describe the datasets and DNN model architectures we used for evaluating \emph{FedLess}. Furthermore, we describe the distribution and configuration of client functions across the different FaaS platforms.

\subsection{Datasets}
\label{sec:datasetsclientdistributions}
\textbf{MNIST}: The MNIST Handwritten Image Database~\cite{lecun-mnist} comprises 70,000 images of handwritten digits with ten classes corresponding to the respective digits. We use all 10,000 test images of the dataset to evaluate the global model after each training round centrally. To simulate a non-IID setting as in \cite{Chadha2020,mcmahan2017communication},
we sort the 60,000 images in the training set by label, split them into 200 shards of 300 images each, and distribute the shards to the clients. This dataset represents an image classification task. 

We also perform experiments on more realistic datasets from the \emph{LEAF} benchmarking framework~\cite{caldas2018leaf}. The datasets chosen are described below and are by nature non-IID. For both datasets, we use the same preprocessing steps as in~\cite{caldas2018leaf}.



\textbf{Shakespeare}: This dataset contains sentences from the \emph{The Complete Works of William Shakespeare}~\cite{shakespeare-gutenberg}, partitioned by the speaker and play. In total, it contains 4,226,158 utterances of length 80 by 1,129 different \emph{users}, with on-average $\sim3,700$ for each. The task of this dataset is to predict the next character in a sentence given the previous 80. 

\textbf{FEMNIST}: It is a modification of the 
EMNIST~\cite{cohen2017emnist} dataset and contains over 800,000 images of digits and characters partitioned by the writer. In total, it contains contributions from 3,550 writers with $\sim226$ images from each on average. Similar to MNIST, the task of this dataset is to classify images. 


In our experiments, we serve all datasets using an \emph{nginx}~\cite{nginx} store with each dataset partition being available at a separate URL and assign it to one client.

\subsection{Model Architectures and Hyperparameters}
\label{sec:modelarchhyperparams}
\textbf{MNIST}: We use the same CNN as in \cite{Chadha2020}, consisting of 2 convolutional layers with kernels of size 5x5, a fully connected layer with 512 neurons, and the output layer with ten neurons. The network uses ReLU as the sole activation function and the categorical-cross-entropy loss function.
In total, the model has 582,026 trainable parameters, which amount to 2.3 MB of memory when serialized. We use a batch size of 10 as in~\cite{Chadha2020} with five local epochs for each client. Because we experienced faster convergence with momentum-based optimization methods, we used Adam as the optimizer.


\textbf{Shakespeare}: Similar to~\cite{caldas2018leaf}, we use an \emph{LSTM}~\cite{hochreiter1997long}. First, the characters in sequences of length 80 are each embedded in an eight-dimensional space, followed by two LSTM layers, each comprising 256 units. In the end, the output layer has the same size as the vocabulary and a softmax activation function. Like~\cite{caldas2018leaf}, we optimize the model using standard Stochastic Gradient Descent with a learning rate of \texttt{0.8} and use the categorical cross-entropy as the loss function. The LSTM has 818,402 trainable parameters, which take $\sim3$MB in serialized form. Due to long training times of \emph{LSTMS}, and limitations with function timeouts in current FaaS providers (\S\ref{sec:bfaas}), we use a batch size of 32 with one local epoch.

\textbf{FEMNIST}: As in~\cite{caldas2018leaf}, we use a large CNN model with two convolutional layers, each followed by a max-pooling layer, followed by a fully-connected layer of size 2048 before resulting in the output layer of size 62. The final model size comes down to 6,603,710 trainable parameters, $\sim26.4$MB when serialized. Similar to MNIST, we use a batch size of 10 with five local epochs and Adam as the optimizer. For both MNIST and FEMNIST, we use the default values for the learning rate.

To calculate a global test accuracy and loss for FEMNIST and Shakespeare, we use the reported test set evaluation metrics from the clients and calculate a weighted average based on the test set cardinalities from the clients (\S\ref{sec:federatedeval},\S\ref{sec:sysworkflow}).


\subsection{Serverless client functions configuration and distribution}
\label{sec:serverlessfuncclusters}
To host our parameter server, we used a virtual machine (VM) with 40vCPUs and 177GB RAM on the LRZ compute cloud~\cite{lrzcc}. On the same VM, we ran an OpenWhisk cluster on top of Kubernetes. All client functions are configured with a memory limit of \texttt{2048}MB, and functions using public cloud providers are deployed in the same region, i.e., \texttt{Frankfurt}, Germany. Only the memory limit for functions on Azure could not be directly configured since it uses dynamic memory allocation with a maximum memory value of 1.5GB per function. We set the memory limit for the aggregation function of \emph{FedLess} to \texttt{4096}MB (\S\ref{sec:sysdesign}).


For all the three datasets (\S\ref{sec:datasetsclientdistributions}) we ran experiments with 200 clients in total and varied the number of clients sampled in each training round. For MNIST and FEMNIST, we deployed 170 Google Cloud Functions (GCF), 10 AWS Lambda functions, 10  IBM Cloud Functions, 5 OpenWhisk functions on an on-premise cluster, and 5 Azure Functions. Training LSTMs using a large number of simultaneous client functions up to convergence is out of our cloud budget. As a result, for the Shakespeare experiments, we only use up to 25 simultaneous clients.

\section{Experimental Results}
\label{sec:results}
We repeat all our experiments five times and follow best practices while presenting our results~\cite{8758926}.

\begin{figure}[t]
    \centering
    \includegraphics[width=0.75\linewidth]{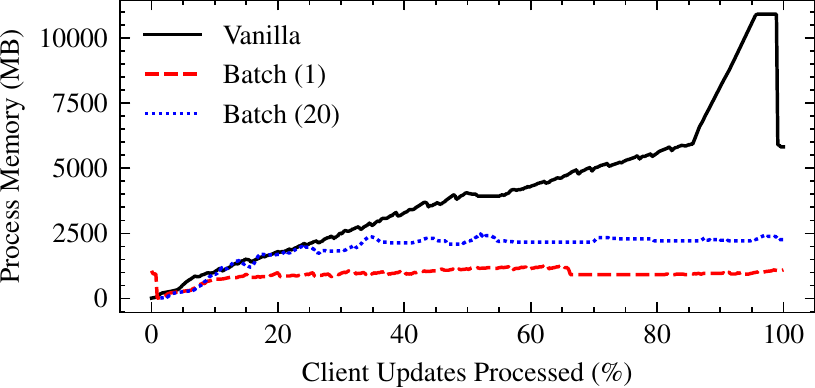}
    \caption{
        Decreased memory footprint of our aggregator implementation.
        Results are generated in a simulation with 200 client updates of the FEMNIST model.
    }
    \label{fig:improved-aggregator}
\end{figure}

\subsection{Analyzing FedLess Performance}
\label{sec:fedless_perf}
As described in \S\ref{sec:rama}, due to the limitations on the amount of memory configurable for a FaaS function, using a standard FaaS function for the aggregation would not be possible for large-scale scenarios. Figure~\ref{fig:improved-aggregator} demonstrates the behaviour of our modified \texttt{FedAvg} implementation in a simulated setting. Aggregating the parameters of 200 clients for the FEMNIST CNN (\S\ref{sec:modelarchhyperparams}) would require more than 6GB of memory for the vanilla \texttt{FedAvg} implementation. In contrast, using batches of 20 updates, our batched running average calculation requires slightly more than 2GB of memory. Note that the overhead created by our implemented security features (\S\ref{sec:gensecfeatures}) is minimal. The actual token validation and checking takes only a few milliseconds in our client functions. Moreover, after a client fetches the token from the Cognito server, we cache it in the function instance's memory, requiring no further requests (\S\ref{sec:bfaas}).





\begin{figure}[t]
    \centering
    \includegraphics[width=1.0\linewidth]{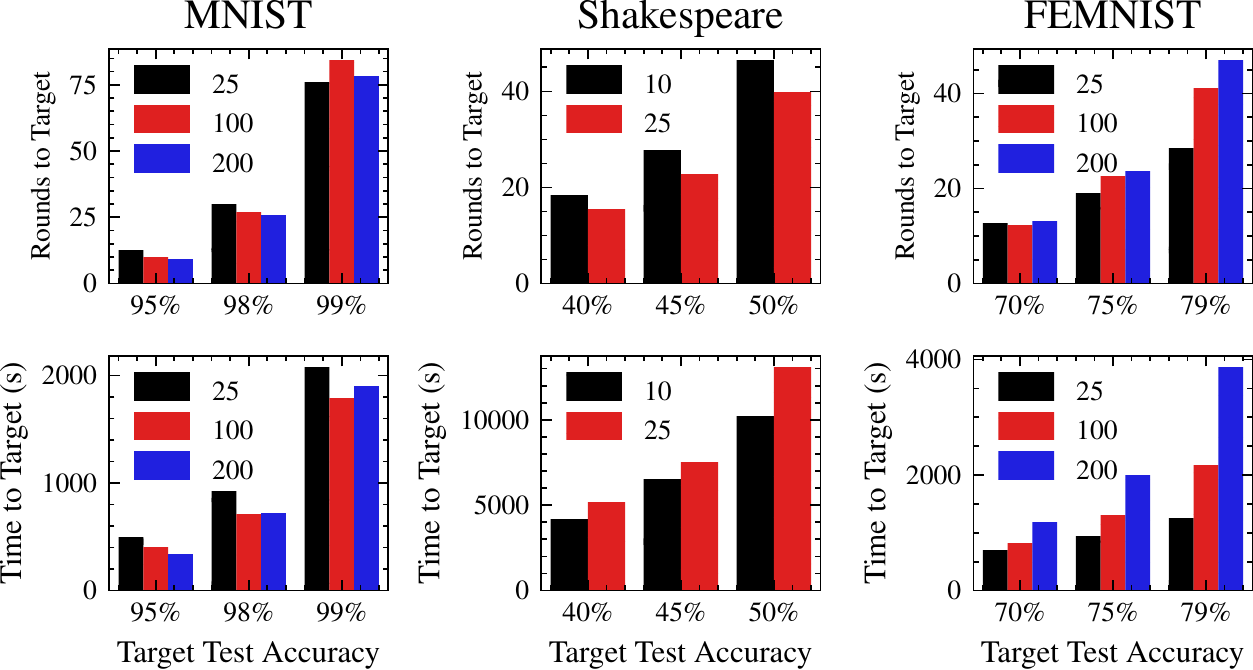}
    \caption{
        Convergence speed of FedLess for different datasets and numbers of clients per round.
        Each dataset is shown in a separate column. 
    }
    \label{fig:fedless-convergence}
\end{figure}

\begin{figure}[t]
    \centering
    \includegraphics[width=0.75\linewidth]{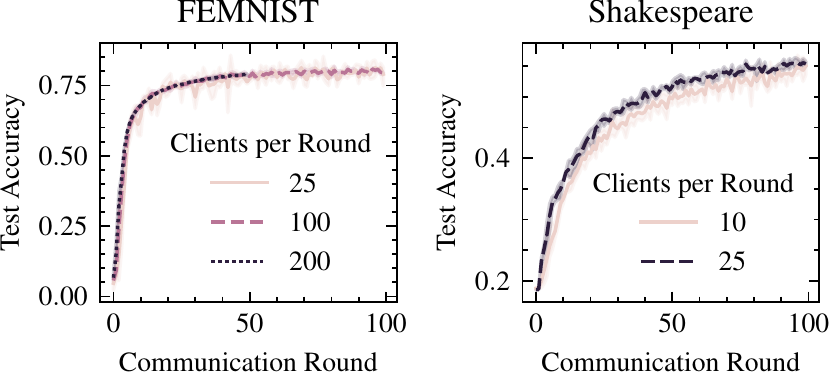}
    \caption{FedLess test accuracies over time for FEMNIST and Shakespeare.}
    \label{fig:fedless-convergence-curves}
\end{figure}

\begin{figure*}[t]
\centering
    \begin{subfigure}{0.3\textwidth}
    \centering
        \includegraphics[width=\columnwidth]{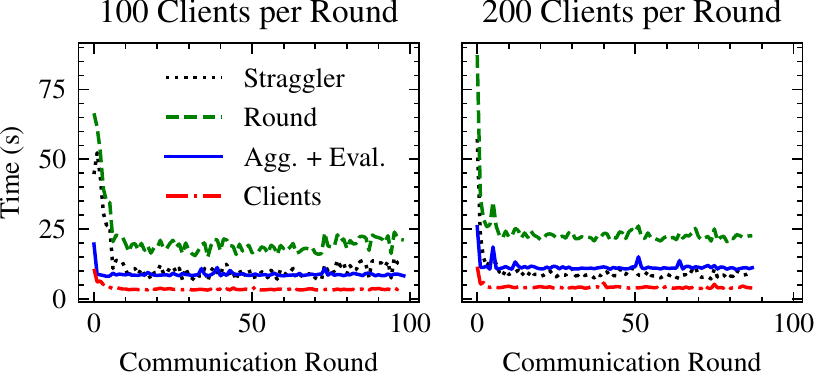}
        \caption{MNIST.}
        \label{fig:mnist-timings}   
    \end{subfigure}
    \begin{subfigure}{0.3\textwidth}
    \centering
        \includegraphics[width=\columnwidth]{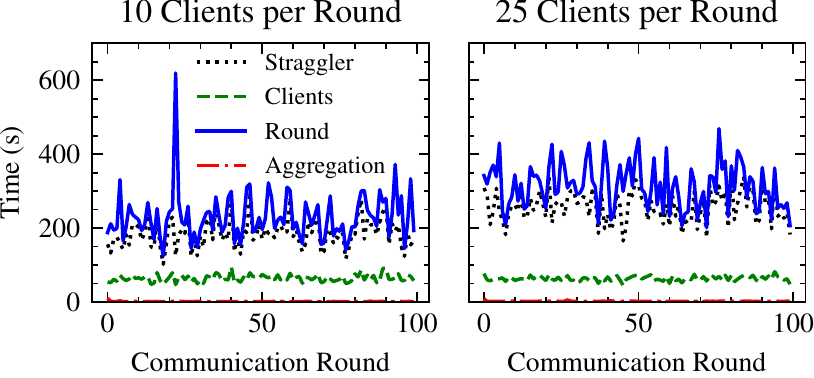}
        \caption{Shakespeare.}
        \label{fig:shakespeare-timings}
    \end{subfigure}
    \begin{subfigure}{0.3\textwidth}
    \centering
        \includegraphics[width=\columnwidth]{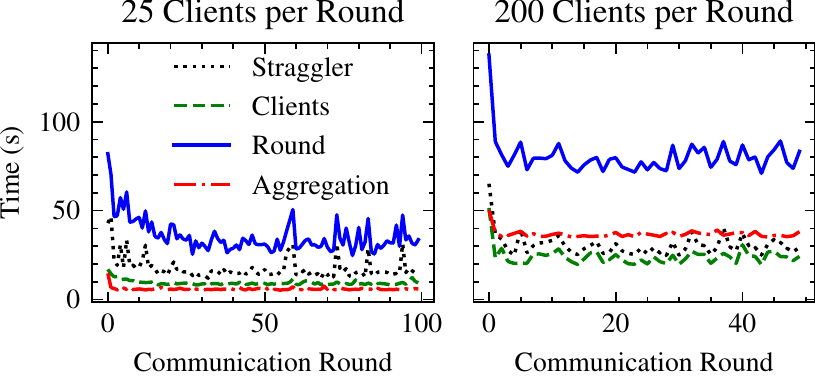}
        \caption{FEMNIST.}
        \label{fig:femnist-timings} 
    \end{subfigure}
    \caption{Mean required time for different steps in FedLess for the different datasets. \emph{Straggler} shows the runtime of the slowest client. If present, \emph{Agg. + Eval.} shows the runtime of the aggregator function that also performs evaluation (\S\ref{sec:sysworkflow}).}
    \label{fig:fedlesstimings}
\end{figure*}

The number of training rounds and the overall time taken by \emph{FedLess} to reach a specific target accuracy for the three datasets is shown in Figure~\ref{fig:fedless-convergence}. Results are shown for different numbers of clients involved in each round. For MNIST, we observe that increasing the number of clients involved in each round did not significantly influence the overall number of required rounds. Using a higher number of clients per round did speed up training in terms of required rounds and time for reaching accuracies less than $99$\%. However, for higher accuracies, the difference is not that evident. Similarly, for Shakespeare, we observe that more clients per round lead to faster convergence only with respect to communication rounds. This behavior is also visible from Figure~\ref{fig:fedless-convergence-curves} and can be attributed to the remarkable variance in the execution time of each client in this dataset due to the non-IID distribution (\S\ref{sec:flback}). As a result, using more clients per round increases the chance of waiting longer for stragglers, increasing the overall time until a certain accuracy is reached. For the FEMNIST dataset, using more clients per round leads to slower convergence in both dimensions, i.e., communication rounds and time to reach the target accuracy. This can be attributed to the large variance in the average reported test set accuracy as shown in Figure~\ref{fig:fedless-convergence-curves}. Using fewer clients in each round for evaluation leads to a larger variance in the average evaluation metrics. 
The overall training behavior for different numbers of clients is approximately the same, showing no significant difference in using more clients per round for our chosen hyperparameters.

\begin{figure}[t]
    \centering
    \includegraphics[width=0.55\linewidth]{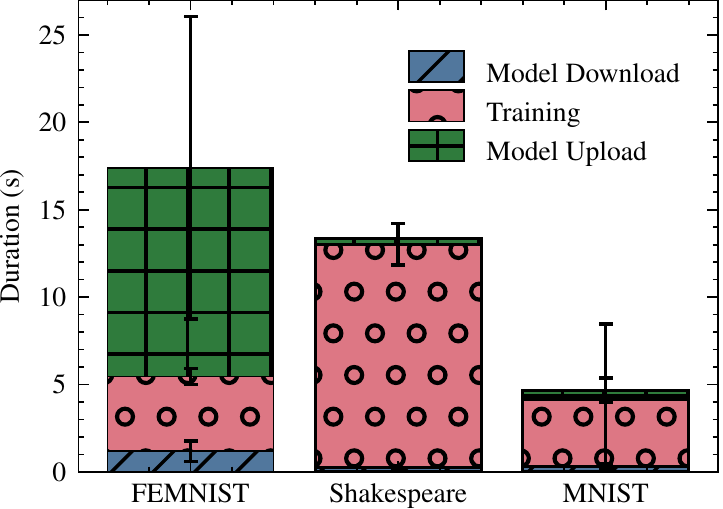}
    \caption{
        Time distributions for FedLess client functions.
        Numbers are based on one client function and multiple training rounds for all three datasets.
    }
    \label{fig:fedless-clients-detailed}
\end{figure}

Figure~\ref{fig:fedlesstimings} shows how long the different actors in our system take over the course of the training for the different datasets. Across all experiments, the total round time is mainly determined by the slowest client of each round, the \emph{straggler}. In each training round, vanilla \texttt{FedAvg} waits for each client to finish or reach a timeout. As a result, only one straggler can drastically increase the total round time. For MNIST, increasing the number of clients from 100 to 200 per round does not impact their execution time, as shown in Figure~\ref{fig:mnist-timings}. The aggregator function becomes slightly slower since it has to load and aggregate results of more clients, slightly increasing the overall round time. Based on these findings, we see that for small local datasets and small model sizes, \emph{FedLess} scales relatively easily. A crucial benefit of our system is that longer round times due to stragglers have almost zero influence on the overall costs since client functions are only billed for their actual runtime and not the time they wait for a new round. Similarly, for the Shakespeare dataset, increasing the number of clients from 10 to 25 did not impact the execution time of the clients, as shown in Figure~\ref{fig:shakespeare-timings}. However, sampling more clients increases the chance of including clients that take a long time to finish. Thus, the overall round time increases. Note that the time required for aggregation or client-side evaluation is negligible for this dataset compared to the time it requires to wait for stragglers. Using an FL strategy that accounts for stragglers~\cite{chen2020asynchronous} is of our interest in the future but out of scope for this work.

For the FEMNIST dataset with the large CNN model (\S\ref{sec:modelarchhyperparams}), increasing the number of clients per round from 100 to 200 almost doubles the total time per round.  Unlike the other two datasets, we observe an increase in the execution time of both the aggregator and the clients. The increase in the aggregator's runtime can be attributed to the networking overhead of loading 200 $\times$ 26MB of parameters and calculating the aggregated model using a running average. The reason for the increase in the mean execution time of the clients can be seen from Figure~\ref{fig:fedless-clients-detailed}.  While the clients training on Shakespeare and MNIST spend most of their time on actual model training and very little on loading or writing parameters from and to the parameter server, FEMNIST clients spend a significant portion of their execution time writing back their results to the parameter server.
Since the parameter download time is much less than the upload time and shows little variability, we can confidently say that the overall networking capabilities of the clients are not the problem. Therefore, the ability of MongoDB's GridFS specification to handle multiple simultaneous uploads of large models is the root cause for the increase in the mean execution time of clients. Investigating the use of another high-performance parameter server with role-based access control rules (\S\ref{sec:parameterserveraccess}) is part of future work. For all datasets, the time for initial rounds is high due to cold starts associated with FaaS functions (\S\ref{sec:bfaas}).

\begin{figure}[t]
    \centering
    \includegraphics[width=1.0\linewidth]{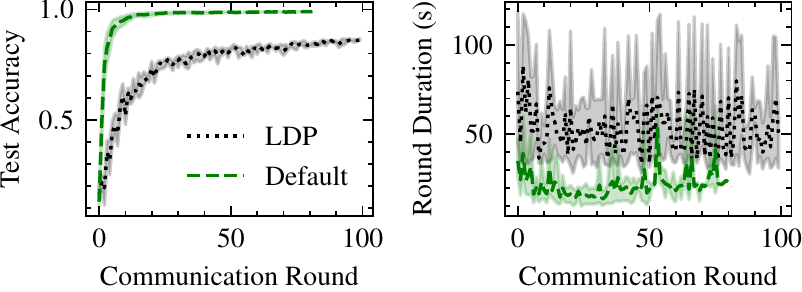}
    \caption{
        Test accuracies and increased round durations for MNIST when using FedLess with LDP across 200 client functions with 25 clients per round.
    }
    \label{fig:ldp-cloud-runs}
\end{figure}

\subsection{Local Differential Privacy}
\label{sec:fedless_ldp}
To demonstrate that it is feasible to use LDP with our system, we train a model on MNIST with 200 client functions (\S\ref{sec:modelarchhyperparams}, \S\ref{sec:serverlessfuncclusters}). We use 25 clients per round, five local epochs, and a batch size of five. After a grid search for the hyperparameters specific to the Adam-DP optimizer in the TF-Privacy library, we use a noise multiplier and L2-clip norm of 1.0, and ten microbatches. The results with LDP as compared to a default MNIST run are shown in Figure~\ref{fig:ldp-cloud-runs}. Due to budget constraints, we do not train the LDP model until convergence but demonstrate that it is possible to use LDP with our system and different supported cloud providers. The significant decrease in test set accuracy reached by our model using LDP in 100 rounds as compared to without can be attributed to the addition of Gaussian noise to the model parameters. However, the same approach might not work for all model sizes directly due to the slower training using DP optimizers. Due to the function execution time limits (\S\ref{sec:bfaas}) on most commercial FaaS providers, training larger models such as LSTMs with LDP will often lead to a function timeout. One could mitigate this problem by only training on a certain number of batches instead of whole epochs or using a smaller microbatch size.


\begin{table}[t]
\centering
\begin{tabular}{lll}
\toprule
{} & \multicolumn{2}{c}{FL Round Duration (s)} \\
 Clients &  FedKeeper &   FedLess \\
\midrule
25  & $18.4\pm9.6$ & $12.3\pm4.6$ \\
50  & $17.9\pm11.5$ & $12.8\pm4.2$ \\
75  & $19.0\pm12.2$ & $14.0\pm3.8$ \\
\bottomrule
\end{tabular}
\normalsize
\caption{FL Round durations of FedKeeper and FedLess for varying numbers of clients.
FedKeeper clients in these experiments already made use of our caching mechanism.
}
\label{tbl:fedkeeper-times}
\end{table}

\begin{figure}[t]
    \centering
    \includegraphics[width=0.50\linewidth]{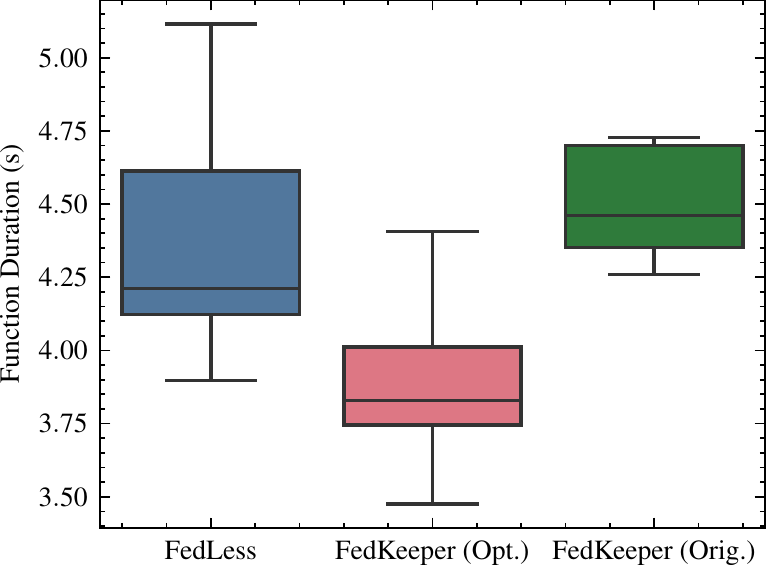}
    \caption{
        Client execution times for FedKeeper and FedLess on MNIST without cold-starts. 
    }
    \label{fig:fedkeeper-client-durations}
\end{figure}



\subsection{Comparison with Fedkeeper}
\label{sec:cmp_fedless_fedkeeper_cmp}
To see how well \emph{FedLess} performs compared to \emph{FedKeeper}~\cite{Chadha2020}, we re-implemented \emph{FedKeeper} with our \emph{FedLess} components. 
Our implementation works almost identical to the original implementation. However, it uses the modular components we implemented for \emph{FedLess}, and we introduced performance improvements like caching. 

We deploy 100 client functions with 2048MB memory for both systems on GCF. 
We only use one cloud provider to minimize the impact of performance variations between FaaS platforms.  We use an on-premise OpenWhisk cluster for the aggregation for both systems and invoker functions for \emph{FedKeeper} (\S\ref{sec:serverlessfuncclusters}). While in \emph{FedLess} the FL client functions are directly invoked by the client invoker in the \emph{FedLess} controller (\S\ref{sec:sysdesign}), \emph{Fedkeeper} uses a separate component called invoker functions which are responsible for invoking the FL clients via an HTTP request. The invoker functions are FaaS functions configured with a memory limit of 256MB. We compare both systems using the MNIST dataset and model (\S\ref{sec:datasetsclientdistributions}, \S\ref{sec:modelarchhyperparams}) for 25, 50, and 75 clients per round.

Table~\ref{tbl:fedkeeper-times} shows the round duration for both the systems. We observe that \emph{FedLess} is always significantly faster than \emph{FedKeeper} with less variability in the round times. The large variance in the round times of \emph{FedKeeper} can be attributed to the cold starts of the invoker functions and their varying runtimes. We re-ran these experiments and exchanged one client function of \emph{FedKeeper} with the original, unoptimized version without caching. The different client execution times are shown in Figure~\ref{fig:fedkeeper-client-durations}. Caching the model architecture and local dataset in-memory of the function instance reduces the median client execution time from 4.46 seconds to 3.83 seconds, a speed-up of roughly 14\%. Only the optimized client functions of \emph{FedKeeper} are slightly faster than the \emph{FedLess} clients, whose median execution time is 4.21 seconds. The additional time in the \emph{FedLess} clients is spent on communicating with the parameter server directly instead of through the invoker functions, which is slightly faster. These results clearly show that \emph{FedLess} is faster and imposes much fewer hardware requirements due to eliminating invoker functions.

\begin{figure}[t]
    \centering
    \includegraphics[width=0.8\columnwidth]{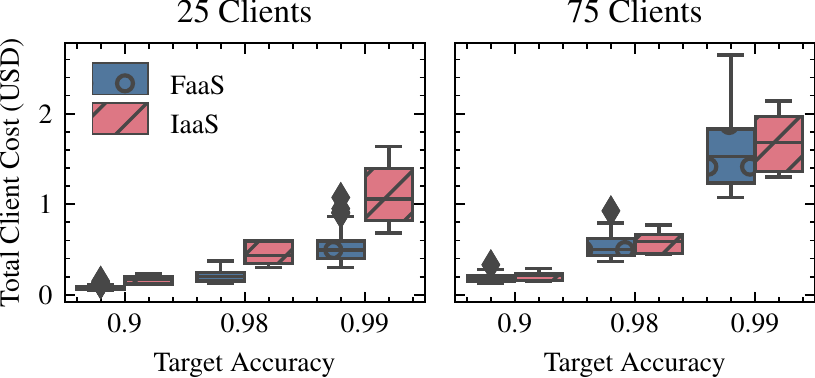}
    \caption{
        Total client costs of FaaS and IaaS to reach a target accuracy on MNIST. 
        Boxplots visualize rough upper and lower bounds depending on the IaaS and FaaS provider's compute capabilities.
    }
    \label{fig:mnist-pricing}
\end{figure}

\subsection{Comparing IaaS and FaaS for FL}
\label{sec:cmp_iaas_faas}
To compare how our system performs compared to a traditional, non-FaaS based FL system, we implemented a baseline system using the open-source FL framework \emph{Flower}~\cite{flower}. We chose \emph{Flower} since it was easy to modify, extend, and integrate into our existing experimentation pipeline. Furthermore, it provides baseline implementations for various FL strategies. For our experiments, we use the default \texttt{FedAvg} implementation. For the communication between the central server and the clients \emph{Flower} uses \texttt{gRPC}. To support invoking a large number of clients in parallel, we had to modify the current \emph{Flower} implementation. For our IaaS based experiments, we use \emph{cgroups} to simulate compute instances similar to  general-purpose instances of public cloud providers (e.g. \emph{m4.large} on AWS EC2~\cite{ec2-instances}, \emph{n2-standard-2} on Google Cloud Compute Engine~\cite{gcloud-instances}) on the LRZ compute cloud~\cite{lrz-cc}.  This is also similar to the setup used in~\cite{Carreira2019} (mostly 2 CPUs, 8GB RAM), and similar to some workloads in~\cite{Jiang2021}. We deploy all Flower clients in separate Docker containers that can each use two vCPUs and 8GB memory. Like in \cite{Jiang2021} our central server for Flower has CPUs and RAM similar to a c5.4xlarge instance on AWS EC2, i.e.,  16 vCPU cores and 32 GB RAM~\cite{ec2-instances}. For a fair comparison, we deploy the client functions for \emph{FedLess} on the same hardware using OpenFaaS. OpenFaaS allows us to limit each client function to 2 vCPU cores and 2GB of memory. As a result, the \emph{FedLess} client functions have the same processing power as the Flower clients. We compare both systems using the MNIST and FEMNIST datasets with 100 clients using the same models and hyperparameters as described in \S\ref{sec:modelarchhyperparams}.

Since both systems use \texttt{FedAvg} and the same hyperparameters, we are primarily interested in timing and pricing differences. Due to the ephemeral, stateless nature of FaaS based systems, we observe that Flower is faster than \emph{FedLess} in all experiments. The relative difference depends on the dataset used. To assess this difference, we use the geometric mean to calculate the mean training round times for \emph{FedLess} and Flower, as proposed in~\cite{hoefler2015scientific}. The mean training round times for \emph{FedLess} are on average $1.77\times$ as long for FEMNIST compared to Flower, whereas they take $1.6\times$ as long for MNIST. However, a key advantage of our system is that client functions that are already finished in a round do not cost anything while waiting and can be scaled to zero. In traditional FL systems, like the one we implemented with Flower, clients simply turn idle while waiting for further requests, accumulating cost, or blocking hardware resources.

\begin{figure}[t]
    \centering
    \includegraphics[width=0.5\columnwidth]{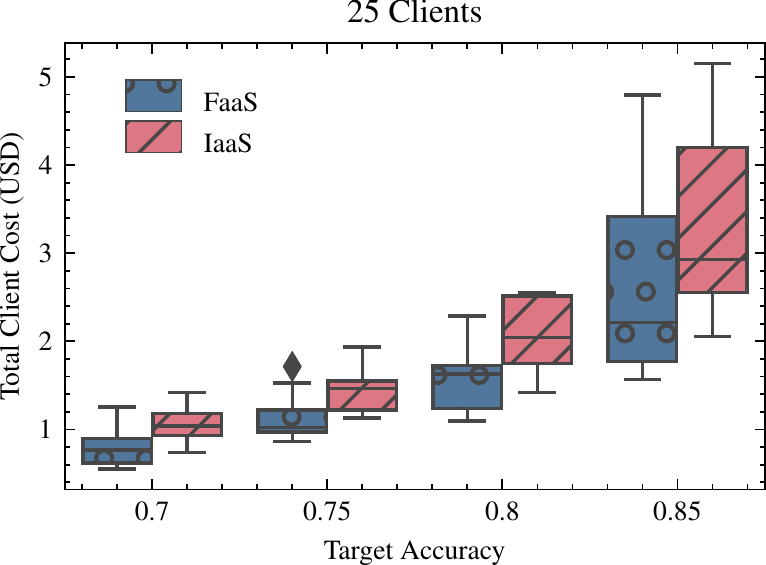}
    \caption{
        Total client costs of FaaS and IaaS to reach a target accuracy on FEMNIST. 
    }
    \label{fig:femnist-pricing}
\end{figure}

To understand the differences between FaaS/IaaS for FL in terms of cost, we created pricing estimates for MNIST and FEMNIST, based on the results of our previous experiments. We base these calculations on the assumption that all clients run on Google Cloud, both for IaaS and FaaS-based training. The compute resources we used for the Flower clients are similar to n2-standard-2 VM instances in both vCPUs and memory~\cite{google-cloud-pricing}. For the \emph{FedLess} clients, we use the execution time presented in experiments from \S\ref{sec:cmp_fedless_fedkeeper_cmp} (MNIST), \S\ref{sec:fedless_perf} (FEMNIST). Since the \emph{Fedless} clients use the same parameter server, hyperparameters, and local datasets, the comparison is fair. For \emph{FedLess} and Flower, we calculate the mean estimated cost of the training runs for both datasets. To not get results skewed by implementation differences, we assume \emph{FedLess} and Flower take an equal number of rounds to reach a particular target accuracy. In addition, for both FaaS functions and IaaS VMs, we ignore additional costs like disk space, costs for cloud buckets, costs to set up and deploy clients, and free-tiers. All prices are calculated for the \texttt{us-central1} region. For our experiments, we only calculate direct costs. For Google Cloud Functions, these are GB-seconds and GHz-seconds~\cite{princinggcf}, invocations and networking costs, and for VMs, they are instance runtime and networking costs.

Since we can only provide estimates, we do not calculate point estimates for the pricing but compute broader bounds. Towards this, we do not only take individual client functions execution time and total training time from the \emph{FedLess}/Flower experiments but additionally include in our calculations how costs would change if the \emph{FedLess} client functions took $2\times$ and $3\times$ as long, and the Flower clients only took $0.5\times$ as long. For the MNIST dataset, the total training costs for both systems until certain target accuracies are reached for different numbers of clients per round is shown in Figure~\ref{fig:mnist-pricing}. We see that overall, the clients' costs are smaller with our FaaS-based approach for the shown settings. Especially further into the training, pricing differences are amplified. However, the cost benefits of using \emph{FedLess} are larger for a smaller number of clients sampled per round. This is intuitive since the more clients are used per round, the less compute time is wasted in the IaaS based approach. Comparing the results for 25 clients per round for MNIST and FEMNIST in Figures~\ref{fig:mnist-pricing} and \ref{fig:femnist-pricing}, we observe that although \emph{FedLess} is still less expensive, the relative difference becomes smaller for larger model sizes. The networking cost due to uploading the updated model is the same for FaaS and IaaS and is responsible for a significant portion of the overall cost. Based on our experiments, we gather first evidence that a FaaS-based FL approach is more cost-efficient for settings where only a fraction of active clients participate in a round. Investigating these trade-offs in detail is part of our future work.

\section{Conclusion and Future Work}
\label{sec:futurework}
In this paper, we presented a novel system and framework for serverless FL called \emph{FedLess}. \emph{FedLess} enables FL across a large fabric of heterogeneous FaaS providers while providing important features such as authentication, authorization, and differential privacy. We demonstrated with comprehensive experiments the features and scalability of our system. In comparison with a traditional IaaS based FL system, we demonstrated the practical viability of our system and showed that, albeit being slower, it can be cheaper and more resource-efficient. 

In the future, we plan to investigate the benefits of switching from our current hybrid system design to a fully event-driven system using message queues and publish-subscribe systems such as Rabbitmq. It also remains to be seen how well our FaaS-based approach works for even larger datasets and models. Although the conventional client heterogeneity in FL is less of a problem in FaaS as compared to traditional IaaS based approaches since idle clients do not cost anything and can scale to zero, we plan to investigate efficient FL using FaaS, which accounts for clients' heterogeneous compute requirements or dataset sizes.

\section{Acknowledgement and Reproducibility}
This work was supported by the funding of the German Federal Ministry of Education and Research (BMBF) in the scope of the Software Campus program. Google Cloud credits in this work were provided by the \textit{Google Cloud Research Credits} program with the award number NH93G06K20KDXH9U.


All code artifacts related to this work are available at\footnote{https://github.com/andreas-grafberger/fedless}.

\bibliographystyle{IEEEtran}
\thispagestyle{empty}
\bibliography{parallelpgm}


\end{document}